\documentclass[%
 reprint,
 amsmath,amssymb,
 aps,
]{revtex4-2}

\usepackage{graphicx}
\usepackage{dcolumn}
\usepackage{bm}
\usepackage{amsmath}
\usepackage{amssymb}
\newcommand{\sech}{\mbox{sech}}

\begin{document}

\preprint{APS/123-QED}

\title{Breather Solutions to the Cubic Whitham Equation}%
\author{Henrik Kalisch}
 \email{Henrik.Kalisch@uib.no}
\affiliation{
Department of Mathematics, University of Bergen, PO Box 7800, 5020 Bergen, Norway}
\author{Miguel A. Alejo}
\affiliation{Departemento de Matem\'aticas, Universidad de Cordoba, Spain}
\author{Ad\'{a}n J. Corcho}%
\affiliation{%
Instituto de Matem\'atica, Universidade Federal do Rio de Janeiro, Brazil}
\author{Didier Pilod}
\affiliation{
Department of Mathematics, University of Bergen, Norway}

\date{\today}

\begin{abstract}
We show numerical evidence that breather solutions exist in the cubic Whitham equation which arises as a water-wave model for interfacial waves. The equation combines nonlinearity with the non-local character of the water-wave problem and is non-integrable as suggested by the inelastic interaction of solitary waves. It generalizes the completely-integrable modified KdV (mKdV) equation. In the mKdV equation, breather solutions appear naturally as ground states of invariant integrals, suggesting that such structures may also exist in non-integrable models, at least in an approximate sense.
\end{abstract}

\maketitle



\section{\label{intro}Introduction}
In this work, we are concerned with the numerical approximation of breather solutions
in nonlinear dispersive equations. A breather is a localized but oscillatory 
wave-packet like solution
of a dispersive equation which is either stationary or propagates in a similar fashion as a solitary wave.
While breathers have been observed experimentally, most if not all theoretical constructions of breathers
are based on {\em explicit solutions} of a {\em completely integrable} nonlinear dispersive differential equation
such as the nonlinear Schr\"{o}dinger (NLS) equation. In the present work, we construct numerical
approximations of breather solutions in a {\em non-integrable} dispersive equation.

Discussions of breather solutions have featured prominently
in the literature on nonlinear dispersive model equations for several decades.
Historically, breather solutions have been considered in completely integrable equations
such as the sine-Gordon equation \cite{seeger1953theorie,ablowitz1973method}
and the modified Korteweg-de Vries (mKdV) equation \cite{wadati1973modified,alejo2013nonlinear}.
So far, the greatest number of exact breather solutions have been found in the
NLS equation. Indeed, the first breather-type solution of the NLS equation
was found over $40$ years ago by Kuznetsov \cite{kuznetsov1977solitons}
and Ma \cite{ma1979perturbed}, followed by the discovery of the
Peregrine breather a few years later \cite{peregrine1983water}.
The authors of \cite{akhmediev1985generation, akhmediev1986modulation}
found a new solution of the NLS equation now called the Akhmediev breather.
This solution is temporally localized,
and for large negative and positive times approaches a plane-wave solution
of the NLS equation. Thus in a sense this solution may be thought of as
the nonlinear stage of the modulational instability observed in
periodic wavetrains in surface water waves \cite{benjamin1967disintegration}
and also exhibited by plane-wave solutions of the NLS equation 
\cite{zakharov1968stability}.

Due to the temporal localization, the Akhmediev breather has been put forward
as a possible mechanism for the development for rogue waves, which are 
unexpectedly large ocean waves that seem to appear out of nowhere.
One common definition of a rogue wave is that the amplitude of the wave
be about two times higher than the surrounding wavefield.
Indeed, one diagnostic for the possible occurrence of rogue waves is
the Benjamin-Feir index \cite{serio2005computation} which essentially measures the probability
of modulational instability to occur. As the Akhmediev breather provides
a path from modulational instability to a singular wave event with large amplitude,
it may be used to predict the occurrence of a rogue wave.

One major step forward in the study of breathers and rogue waves was the
observation of the Akhmediev breather and associated large wave event
in an experimental wave flume \cite{chabchoub2011rogue}.
Subsequently, the authors of \cite{chabchoub2012super}
used an ingenious method to use a $9$ meter wave tank for the creation
of a higher-order breather by restarting the experiment seven times
using data from the previous experiment as starting data for the next run.
The authors also provided convincing comparisons between experimentally obtained wave profiles 
and the exact breather solutions of the NLS equations to show that the Akhmediev breather
had indeed been created in the laboratory.

One feature that all of the theoretical breather solutions mentioned above
have in common is that they are {\em exact} or {\em closed-form} solutions
of a {\em completely integrable} partial differential equation. On the other hand,
as shown in the experiments in \cite{chabchoub2011rogue,chabchoub2012super}
breathers exist in reality, and there is no reason one would expect that
breather solutions only exist in completely integrable models.
Approximate breather solutions were also considered in internal wave models.
Indeed, the fully-integrable Gardner equation 
features exact breather solutions \cite{pelinovsky1997structural},
but it is shown in \cite{lamb2007breather} that
when inserted into non-integrable models,
the solutions are constantly radiating energy which casts doubt on
whether these are truly coherent structures in the non-integrable models.
In \cite{slunyaev2013super},
approximate breather solutions are observed in
two non-integrable models: the Dysthe equation \cite{dysthe1979note}, essentially
a higher-order NLS equation, and the full water-wave problem based
on the inviscid and incompressible Euler equations.
However, as can be seen in figures 5 and 6 in \cite{slunyaev2013super}
these numerical solutions are not as regular as the closed-form
solutions of the NLS equation such as the Akhmediev breather.
Indeed, a slight asymmetry is observed in the solutions provided
in \cite{slunyaev2013super}, and the solutions are also radiating energy.
This may be due to the non-integrability
of the models used in \cite{slunyaev2013super}
and may also be connected
with wave breaking which was observed in the experiments reported
in \cite{slunyaev2013super}.

As suggested by the short summary above, the study of breather solutions
is an active field of research with a wide range of open problems 
concerning existence, stability and relation to extreme events.
For the NLS equation, new breather solutions are found in closed form on a regular basis
(see for example \cite{gaillard2015peregrine}), and there are several results showing
existence \cite{alejo2013nonlinear}, stability \cite{haragus2021linear,alejo2021stability},
or non-existence or instability  \cite{denzler1993nonpersistence,munoz2019breathers,munoz2017instability}.
In the present work we present numerical evidence for the existence of breather solutions
in a non-integrable model.

\section{The cubic Whitham equation}
The Korteweg-de Vries (KdV) equation 
is a simplified model equation for waves at the surface of an inviscid
incompressible fluid. The equation includes the essential effects of 
nonlinearity and dispersion, and balancing these two effects is the basic mechanism behind the
existence of both solitary-wave solutions and periodic traveling waves. 

The KdV equation can also be used as a model for waves at the interface in a two-fluid system
although it has been noted that the so-called extended KdV (eKdV) equation
also known as the Gardner equation is more advantageous
for internal waves as it more closely describes the typically broader
solitary waves \cite{stanton1998observations}. 
In the two-layer case, the unknown function $\eta(x,t)$ 
is the deflection of the interface from its rest position.
\begin{figure}[b]
\centering
\includegraphics[width=0.48\textwidth]{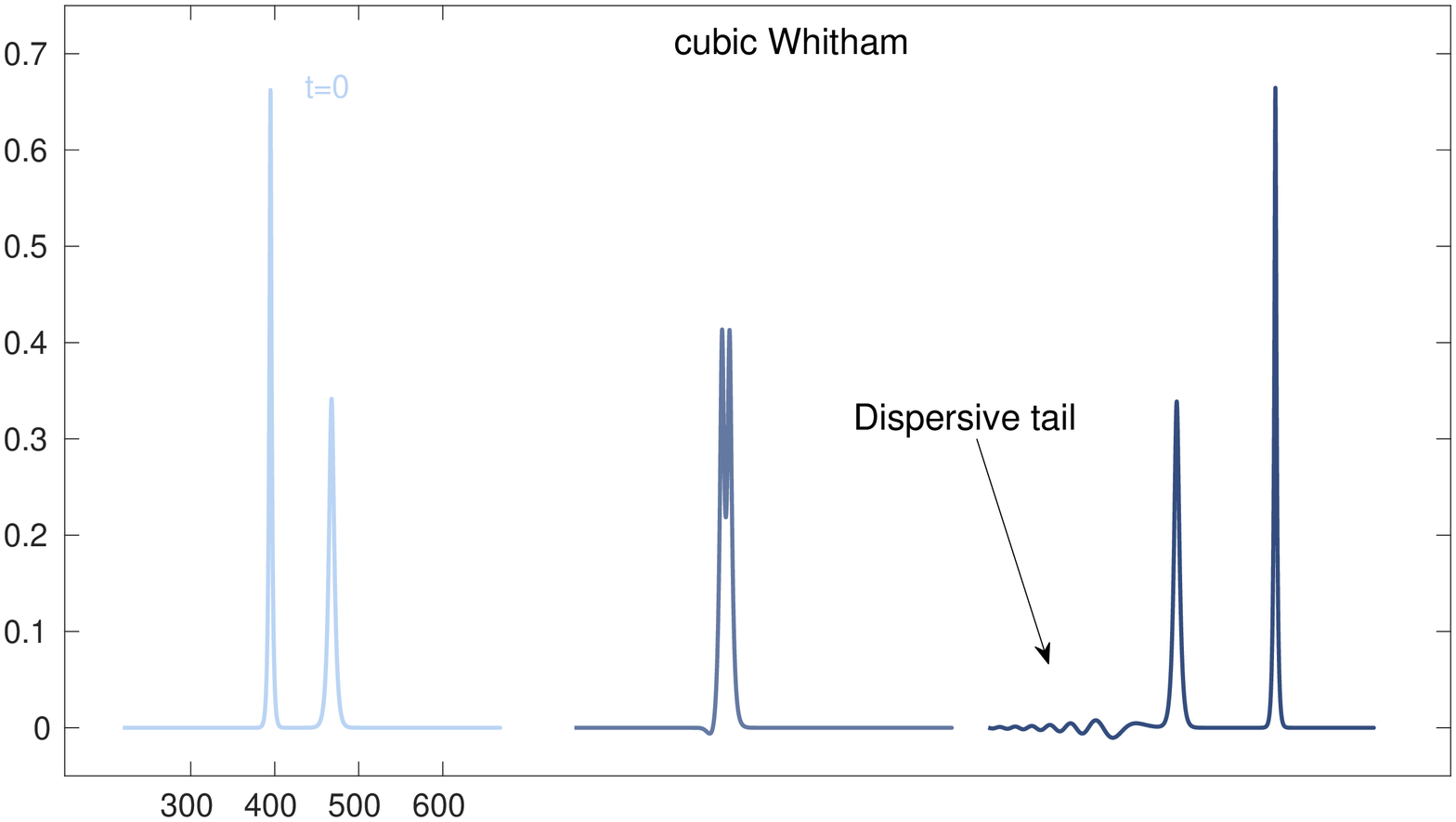}
\includegraphics[width=0.48\textwidth]{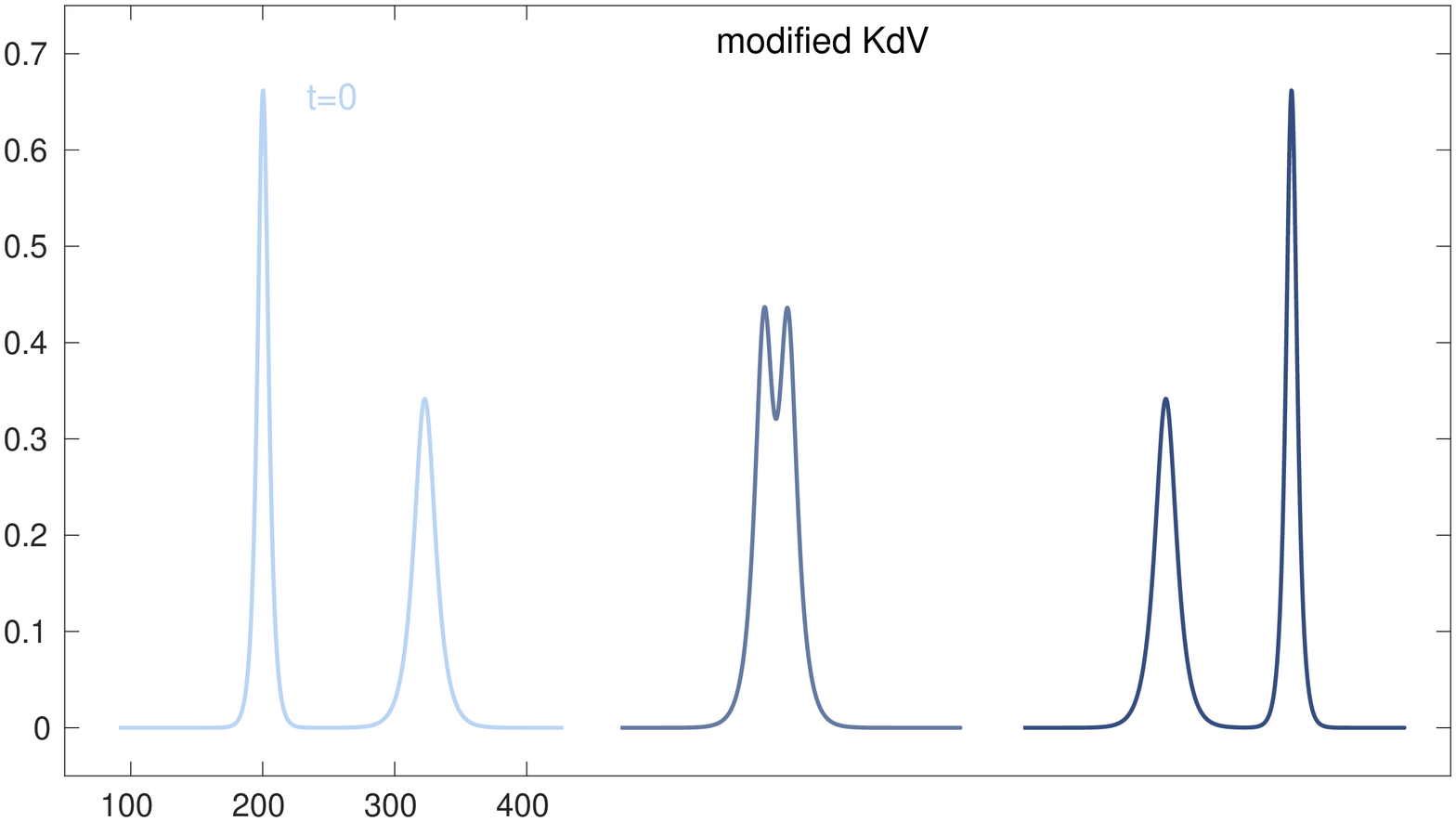}
\caption{\small Interaction of solitary waves of amplitude $0.67$ and $0.34$
in the the cubic Whitham equation \eqref{cWh} (top panel) 
and in the modified KdV (mKdV) equation \eqref{mKdV} (lower panel). 
Solitary waves are propagating to the right.
Pre-interaction (left), interaction point (center) and post-interaction (right) are shown.
A dispersive tail is visible in the post-interaction plot
in the case of the cubic Whitham equation, but not in the mKdV equation. 
Note that the position of the pre-and post-interaction plots on the $x$-axis is not to scale.
Nevertheless, it can be seen that solitary waves of the 
same height are much wider in the mKdV equation than in the cubic Whitham equation.}
\label{Fig2:Interaction}
\label{Fig1:Interaction}
\end{figure}

Following an idea of G.B. Whitham \cite{whitham1965general},
one may use the full representation of the the dispersion relation for a two-fluid system.
In particular for a lower layer of depth $h_1$ and density $\rho_1$
and an upper layer of depth $h_2$ and density $\rho_2$, the equation
will have the form
\begin{equation}\label{ekdvWhitham}
\eta_t + \alpha_1 \eta \eta_x  + \alpha_2 \eta^2 \eta_x +  K_I *  \eta_{x} = 0,
\end{equation}
where the coefficients are given in terms of
$c_0^2 =  g h_1h_2 \frac{\rho_1-\rho_2}{\rho_1 h_2 + \rho_2 h_1}$
by
$\alpha_1 = \frac{3}{2} c_0 
                 \frac{1}{h_1h_2}
                 \frac{\rho_1 h_2^2 - \rho_2h_1^2}{\rho_1 h_2 + \rho_2 h_1}$
and
$\alpha_2 = \frac{3c_0}{(h_1h_2)^2}
                  \Big[\frac{7}{8}\Big(\frac{\rho_1 h_2^2 - \rho_2 h_1^2}{\rho_1 h_2 + \rho_2 h_1}\Big)^2
                - \frac{\rho_1 h_2^3 + \rho_2 h_1^3}{\rho_1 h_2 + \rho_2 h_1}\Big]$.
The convolution operator is given by the kernel 
\begin{equation}\label{eq:KI}
K_I = \mathcal{F}^{-1} 
\left(\sqrt{\textstyle{ \frac{g (\rho_1-\rho_2) \tanh (k h_1) \tanh (k h_2)}
                             {\rho_1 \tanh(k h_2) + \rho_2 \tanh(k h_1)}
                      }} 
\right),
\end{equation}
where $\mathcal{F}^{-1}$ is the inverse Fourier transform, defined by
\begin{equation*}
  \mathcal{F}^{-1}U (x)  = \frac{1}{2 \pi} \int_{-\infty}^\infty U(k) e^{ikx} \, \mathrm{d}k.
\end{equation*}
(see \cite{koop1981investigation, stanton1998observations, 
craig2005hamiltonian, kalisch2010stability}).
In the case when $\rho_1 h_2^2 \sim \rho_2 h_1^2$, the quadratic
coefficient vanishes, and the pure cubic equation
(without the quadratic term) appears as the most appropriate model.
The equation can be normalized by taking the depth of the lower
fluid $h_1$ as the unit of distance, and $h_1/c_0$ as a unit of time.

For the purpose of this article, we use the the so-called \emph{cubic Whitham equation}, 
which is a simplified version of \eqref{ekdvWhitham} given by
\begin{equation}
\label{cWh}
\eta_t + \eta^2\eta_x + K \ast \eta_x = 0
\end{equation}
with
\begin{equation}\label{eq:K}
K = \mathcal{F}^{-1} \sqrt{\textstyle{ \frac{\tanh k}{k}}}.
\end{equation}
Equation \eqref{cWh} has the advantage of being closer to the original quadratic Whitham equation
which has seen great attention lately
(see \cite{moldabayev2015whitham,klein2018whitham,carter2018bidirectional,kharif2018nonlinear} 
and references therein), and it can also be more easily compared with the modified KdV equation
\begin{equation}
\label{mKdV}
\eta_t + \eta^2\eta_x + \eta_{xxx} = 0.
\end{equation}
\begin{figure}[t]
\centering
\includegraphics[width=0.5\textwidth]{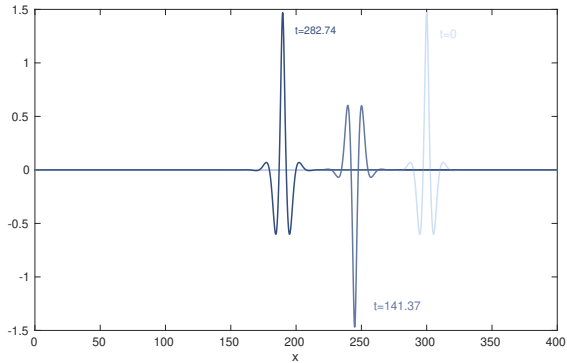}
\caption{\small Evolution of a breather in the mKdV equation
\eqref{mKdV}. The breather propagates to the left.
The right-most profile is taken the initial data, a breather with
$\alpha = 0.3$ and $\beta = 0.4$.
The wave then evolves in the mKdV equation, 
forming a negative shape (center waveform), then evolving
further until it reaches the original shape, but translated left.
}
\label{Fig2:mKdVBreather}
\end{figure}
It is well known that the mKdV equation is an integrable model equation,
and solutions may be found with the help of the inverse scattering transform method,
or using the Miura transform which maps solutions of the KdV equation into
solutions of the mKdV equation \cite{miura1968korteweg}.
On the other hand, the cubic Whitham equation \eqref{cWh} is not integrable as can be seen
explicitly by examining the interaction of two solitary waves.
The existence of traveling-wave solutions and the existence and stability of solitary wave solutions
to the cubic Whitham equation \eqref{cWh} was studied in
\cite{ehrnstrom2009traveling,ehrnstrom2012existence,ehrnstrom2013global,
sanford2014stability, stefanov2020small}.
More recently the formation of shocks has been proved in \cite{klein2020modified}.
Approximate solitary-wave solutions can be found numerically using a
manual cleaning process such as explained in \cite{bona2000models}.
As shown in Figure \ref{Fig1:Interaction}, the interaction between
two solitary waves in the cubic Whitham equation creates a small
dispersive tail after the interaction, while the same code applied
to two solitary waves of the mKdV equation shows a clean interaction.
The appearance of small dispersive ripples after the interaction is
usually a strong indicator of non-integrability.

The cubic Whitham equation \eqref{cWh} admits the following conserved quantities:
\begin{subequations}
\begin{equation}
\mathcal{I}_1=\int \eta \, dx,
\end{equation}
\begin{equation}
\mathcal{I}_2=\int \eta^2 \, dx, 
\end{equation}
\vskip 0.2in
\begin{equation}
\mathcal{I}_3=\int  \Big( \eta \, K_1 \! \! \ast \!  \eta + \frac{1}{4} \eta^4 \Big) \, dx,
\end{equation}
\end{subequations}
where the integration is over the real line, or over one wavelength depending on the situation.
Since the equation is non-integrable, it is unlikely that further conserved integrals can be found. \\

\section{Breather solutions}
One of the most prominent examples of a breather solution is the Peregrine breather 
which is an exact solution of the 
NLS equation \cite{peregrine1983water}. The Peregrine breather can be obtained
by perturbing the exponential Stokes-wave solution of the NLS equation 
with specially chosen functions of polynomial decay.

In most of the models mentioned above, in particular the mKdV and Gardner equations,
breather solutions share common features such as having two free parameters
which can be interpreted as the amplitude and the rate of the inner oscillation of the wave packet.
In particular, the mKdV equation \eqref{mKdV} has a closed-form breather solution given 
in terms of the {\em phase velocity} $\gamma = 3 \alpha^2 - \beta^2$ and the
{\em crest velocity} $\delta = \alpha^2- 3 \beta^2$.
The evolution of a typical mKdV breather is shown in Figure \ref{Fig2:mKdVBreather}.
The formula is represented as
\begin{widetext}
\begin{equation}
\eta(x,t) = 2 \sqrt{6} \beta \sech{\beta(x+\gamma t)}
\Big[
\frac{\cos(\alpha(x+\delta t)) - (\beta/ \alpha) \sin(\alpha(x-\delta t) ) \tanh(\beta ((x+ \gamma t))}{ 1+ (\beta/\alpha)^2 \sin^2(\alpha(x+\delta t)) \sech^2(\beta (x+ \gamma t))}
\Big].
\end{equation}
\end{widetext}
\begin{figure}[t]
\centering
\includegraphics[width=0.5\textwidth]{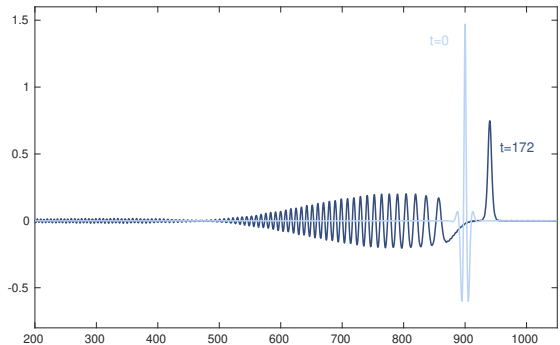}
\caption{\small Evolution of a breather in the quadratic (proper) Whitham equation
$\eta_t + \eta \eta_x + K \ast \eta_x = 0$.
The light blue curve shows the initial data. The dark blue curve shows
the numerical solution at $t=172$.
As in the previous figure, the initial data is given by a breather solution for
the mKdV equation with $\alpha = 0.3$ and $\beta = 0.4$.
The solution develops into a positive solitary wave, 
a nonlinear dispersive structure similar to a dispersive shock wave,
and a linear dispersive tail.}
\label{Fig3: ResolutionWhitham}
\end{figure}
\begin{figure}[]
\centering
\includegraphics[width=0.5\textwidth]{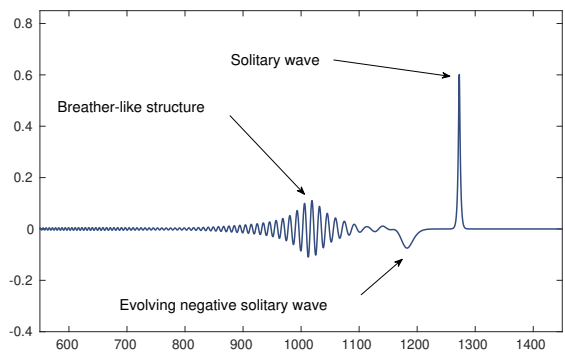}
\caption{\small Appearance of a breather in the cubic Whitham equation \eqref{cWh}.
In this case, the initial data is given by a slightly perturbed breather solution for
the mKdV equation with $\alpha = 0.3$ and $\beta = 0.4$.
The solution develops into a positive solitary wave, an emerging negative solitary wave,
an oscillatory breather-like structure and dispersive noise, shown at $t=1750$. 
A large domain is necessary in order to allow the various structures to separate.}
\label{Fig4:ResolutionModifiedWhitham}
\end{figure}

However, in recent work, breathers have been characterized as solutions of a fourth-order 
ordinary differential equations which implies that breather solutions are critical points 
of a second-order Lyapunov functional, composed of a linear 
combination of low order conserved quantities \cite{alejo2013nonlinear, alejo2015dynamics, alejo2017variational}.
This work essentially raises the question whether breathers exist more generally as stable structures
in non-integrable evolution equations.

For the cubic Whitham equation \eqref{cWh} (as for the quadratic Whitham equation) no closed-form
solutions are known. In particular, it is not known whether \eqref{cWh} features breather-type
solutions. Here, we present numerical simulations which strongly suggest that \eqref{cWh}
does admit breather solutions. In order to find a breather, one may consider the evolution
of initial data in the form of a Gaussian, or possibly a modulated Gaussian. As the solution
develops, in many cases, solitary waves and breathers appear naturally.

For the purpose of approximating solutions of \eqref{cWh},
a Fourier-collocation method is used. This method has the advantage
that the symbol of the linear operator appearing in \eqref{cWh} can
be represented exactly. The discretization and numerical implementation
are similar to methods used in \cite{ehrnstrom2013global}.
The code used here is versatile enough so that it can be used
for the KdV, Whitham and mKdV and cubic Whitham equations
by simply changing coefficients.

In order to find coherent structures for the Whitham equation,
we insert an mKdV breather solution as initial data
into the fully discrete scheme.
As a test, we first solve the mKdV equation itself and
observe that the breather propagates as it should (see Figure \ref{Fig2:mKdVBreather}).
Since this solution has an exact form, this procedure was also
used to test the implementation of the numerical code.

We then insert a perturbed breather into the quadratic and cubic Whitham equation.
In the quadratic Whitham equation, a solitary wave and a dispersive shock wave are formed
(see Figure \ref{Fig3: ResolutionWhitham}).
In the cubic Whitham equation, the initial data evolve into a positive solitary wave, 
a negative solitary wave, a breather-like structure and a dispersive tail
(see Figure \ref{Fig4:ResolutionModifiedWhitham}).
\begin{figure}[h]
\centering
\includegraphics[width=0.5\textwidth]{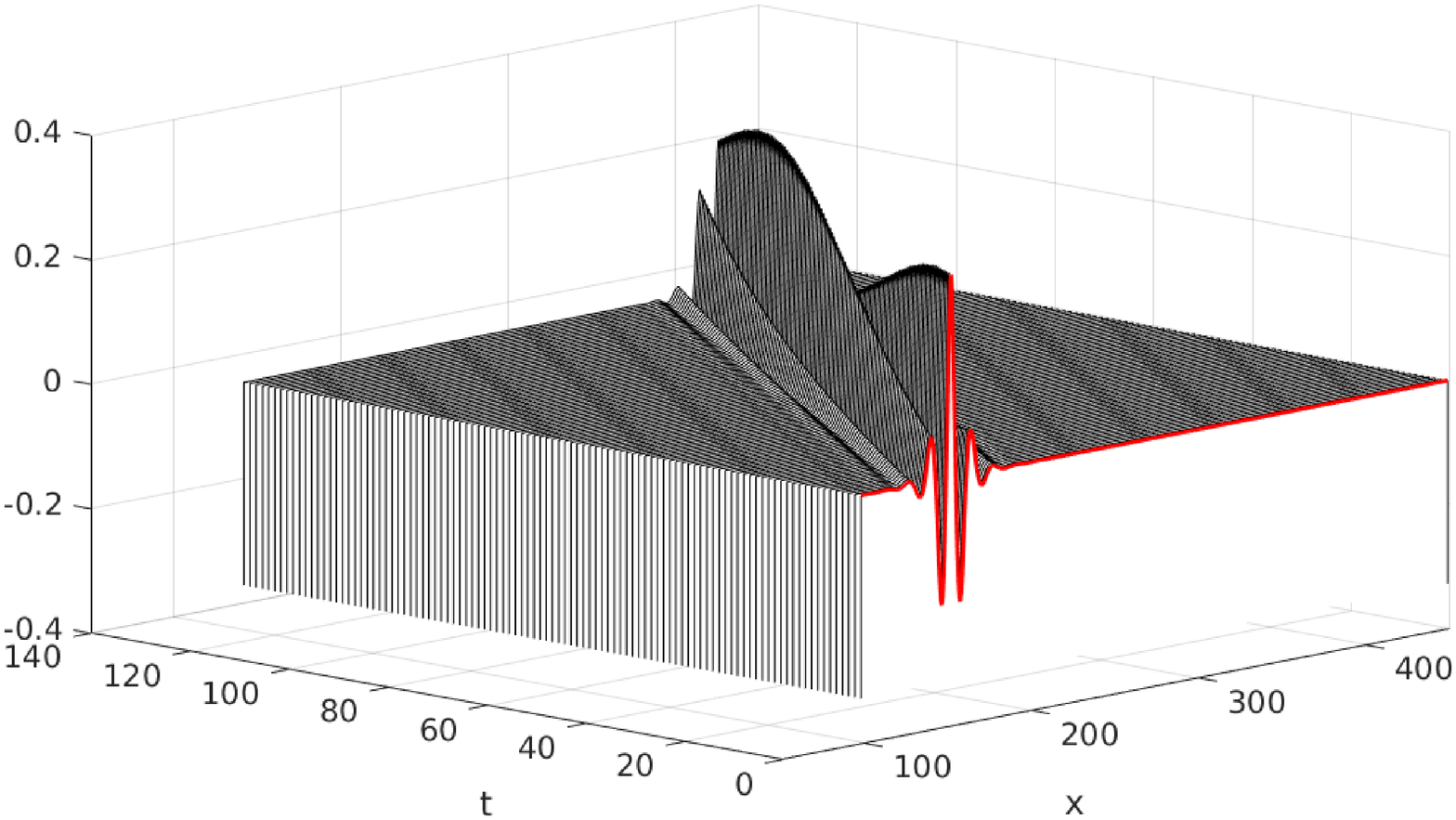}
\includegraphics[width=0.5\textwidth]{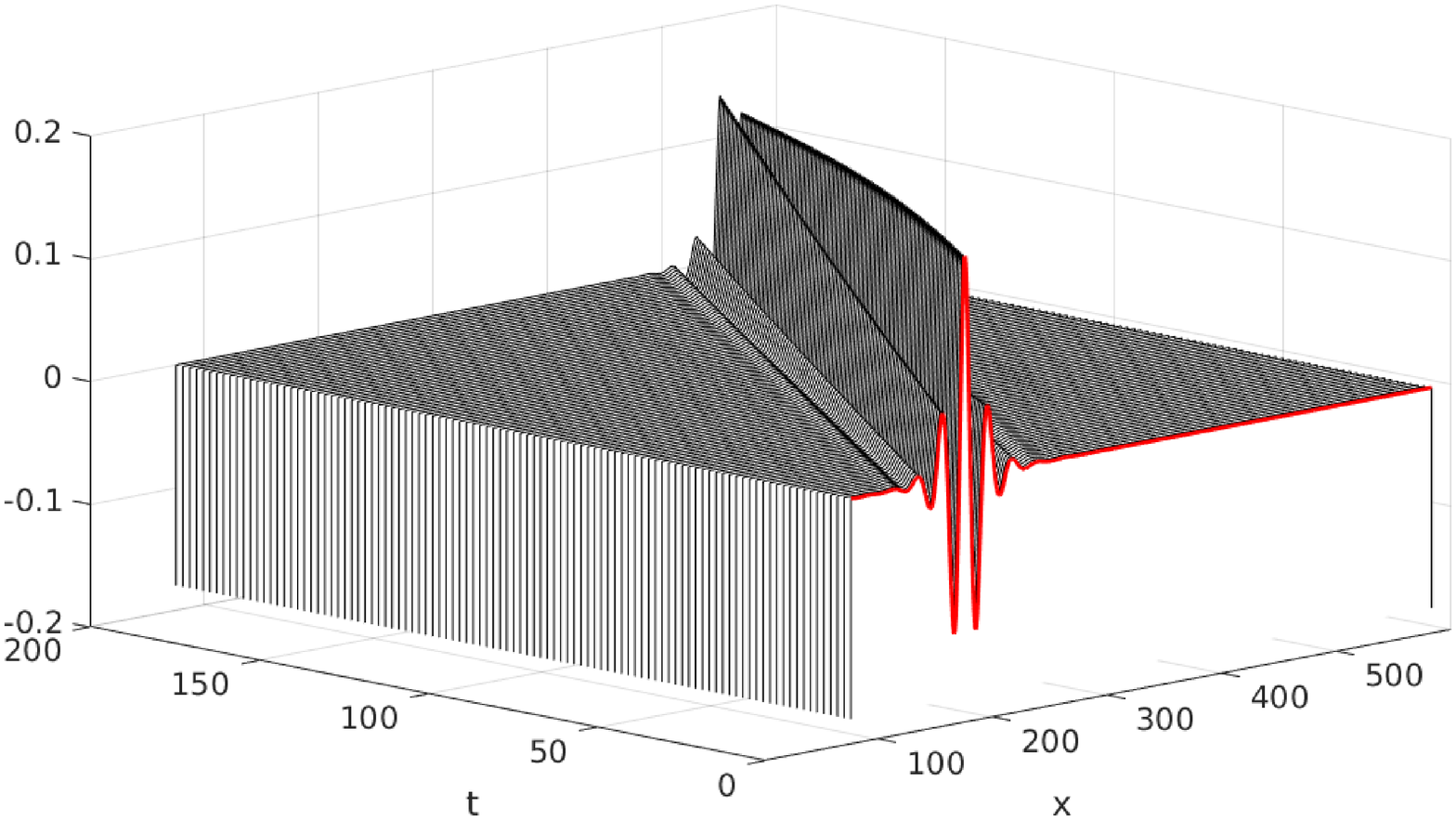}
\includegraphics[width=0.5\textwidth]{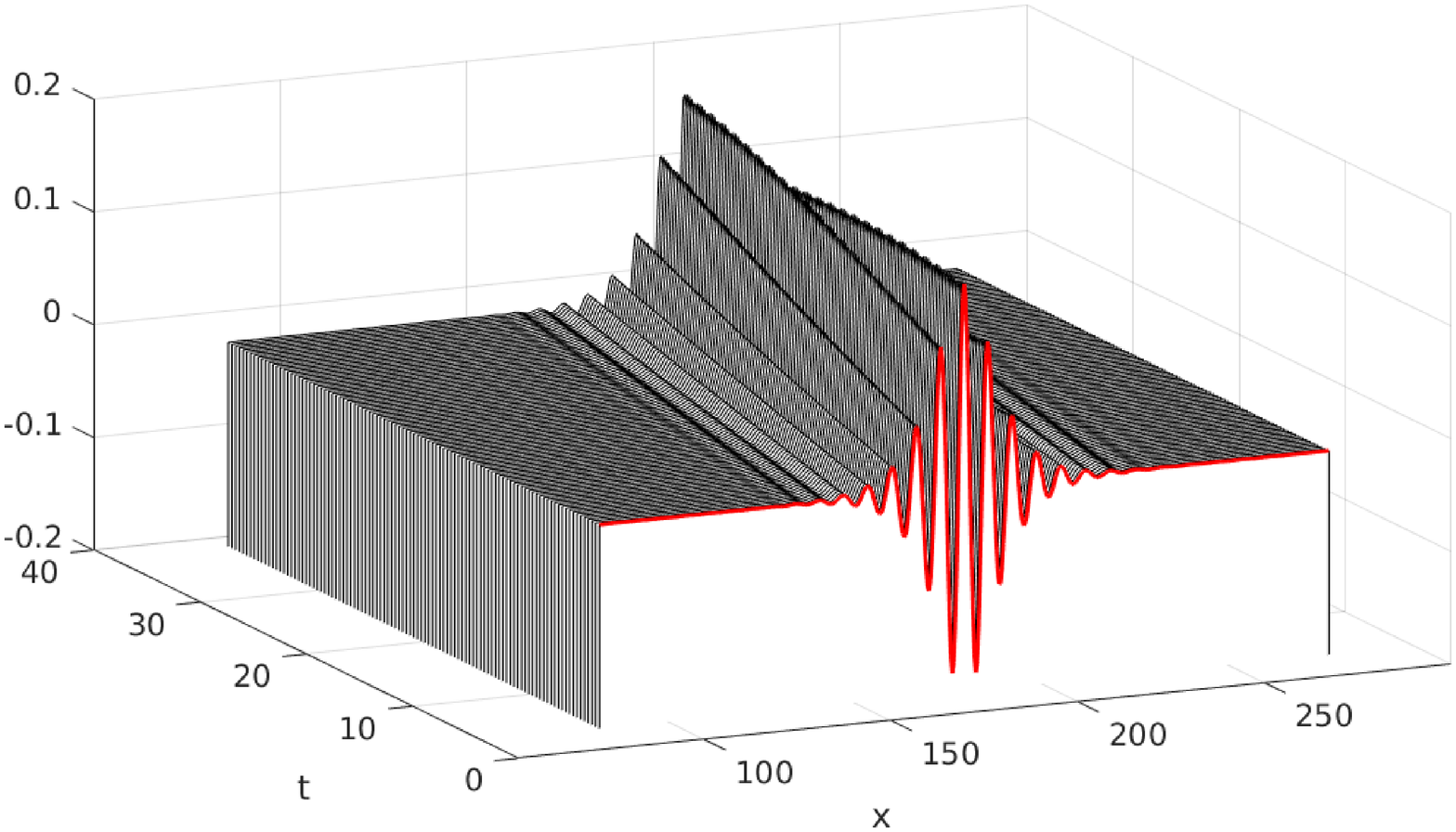}
\caption{\small Time development of breathers in the cubic Whitham equation \eqref{cWh}. 
The breather in the upper panel has a (highest) crest to (lowest) trough waveheight of $0.53$ and
a non-dimensional speed of $0.9$.
The breather in the middle panel has a crest-trough waveheight of $0.30$ 
and a non-dimensional speed of $0.96$.
The breather in the lower panel has a crest-trough waveheight of $0.34$ 
and a non-dimensional speed of $0.69$.
}
\label{Fig5:Breathers}
\end{figure}

\begin{figure}[]
\centering
\includegraphics[width=0.45\textwidth]{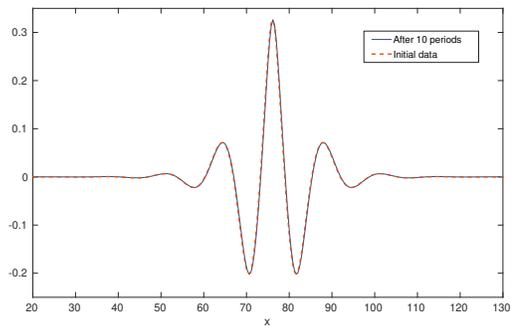}
\caption{\small Evolution of a breather in the cubic Whitham equation \eqref{cWh}
for $10$ wave periods. The red dashed curve is the initial profile of the
approximate breather, shifted on the $x$-axis by $10 \cdot T \cdot c$, where $T$ is the period
and $c$ is the phase speed. The blue solid curve is the
profile after integrating numerically for $10$ periods.}
\label{Fig6:Evolution}
\end{figure}

In order to extract numerical approximations to a breather, a cleaning approach
needs to be used such as explained in \cite{bona2000models}). 
In other words, at some intervals, the dispersive ripples are
simply zeroed out by hand, and the code is restarted. If the coherent structure
(solitary wave of breather) is stable, and the small oscillations are propagating
at a different speed than the main structure, then this process eventually leads
to a close numerical approximation of the target structure, with small ripples
essentially absent almost to machine precision.
The time evolution of three different breathers in the cubic Whitham equation \eqref{cWh}
is depicted in Figure \ref{Fig5:Breathers}.
Figure \ref{Fig6:Evolution} shows the first breather from Figure \ref{Fig5:Breathers}
after evolving for $10$ periods, and no difference
can be discerned visually between the starting waveform and the evolved wave.

\section{Discussion}
The three breathers shown in Figure \ref{Fig5:Breathers} suggest that a rich variety
of different breathers exist in the cubic Whitham equation \eqref{cWh}.
Indeed, comparing the breather shown in the upper and middle panel, the breather
shown in the middle panel appears slightly more oscillatory, but has a much larger
total waveheight (defined here as the elevation of the highest crest minus the
elevation of lowest trough in the wave packet) than the breather in the upper panel.
On the other hand, the breather in the lowest panel has a similar waveheight
to the one in the middle panel, but features many more oscillations.
This suggests that there is probably a two-parameter family of breather
solutions, in analogy with the mKdV equation.
Since no radiation was detected with the numerical code used here,
our numerical results suggest the mathematical existence
of true breather solutions.

The question of the existence of breather solutions for non-integrable equations
was raised in \cite{kosevich1974selflocalization, dashen1975particle}.
These authors provided an asymptotic development suggesting the existence
of a breather solution in the so-called Klein-Gordon $\phi^4$ equation.
However, it was ultimately shown that this expansion did not converge,
and the purported breather solution actually 
features a tiny amount of radiation, though the rate of radiation
is so small that it can only be detected using asymptotics beyond all orders
\cite{segur1987nonexistence,soffer1999resonances}.
The solutions constructed here are also different from the sine-Gordon breather
in the sense that they are propagating, and have a well defined phase and crest velocity,
similar to the two-parameter family of mKdV breathers. In contrast, the sine-Gordon breather 
is a one-parameter family of stationary breathers.

Since the present work is numerical, it is possible that the breathers found here also feature
very small radiation which may be below machine precision,
and in this case the breathers found here would also be approximate in the mathematical
sense, i.e. decaying after a very long time.
On the other hand, the question whether the breather is an exact solution
or a slowly radiating meta-stable state may not be of much importance
in practice. Indeed while it has been shown for example that the Peregrine breather
is unstable to virtually all perturbations \cite{klein2017numerical,munoz2017instability},
breathers are nevertheless observable in the laboratory \cite{chabchoub2012spectral}.
Moreover asymptotic model equations such as the KdV and Whitham equations are generally valid
physical models only on a time scale inversely proportional to the amplitude
\cite{lannes2013water, klein2018whitham, saut2021long, emerald2021rigorous}. 

We should also mention that there are some works discussing mathematical existence
of structures more similar to the sine-Gordon breathers.
In \cite{blank2011breather}, \cite{mandel2021variational} and \cite{hirsch2019real} 
non-explicit real-valued, time-periodic and spatially localized solutions are constructed 
for nonlinear non-integrable wave equations 
with periodic potentials using variational methods 
and spatial dynamics and center manifold reductions.
These solutions are of breather type, but are stationary, and not propagating,
so they are more similar to the breather of the sine-Gordon equation.
It will be interesting to see whether
breather solutions are possible in other fully dispersive equations
such as the fully dispersive Gardner equation \cite{carter2021cubic}
and the fully dispersive NLS equation \cite{trulsen2000weakly}.

\begin{acknowledgments}
This research was initiated during several visits of M.A. Alejo and
A.J. Corcho to Bergen in 2018 and 2020, and M. Alejo and A.J. Corcho
would like to thank the University of Bergen for kind hospitality.
The authors would like to thank Vincent Duch\^{e}ne and Claudio Mu\~{n}oz
for helpful discussions.
This research is supported by the Trond Mohn Foundation
and by the Research Council of Norway under grant number 239033/F20.
\end{acknowledgments}

\bibliography{KACPbib}

\end{document}